\documentclass[epj,nopacs]{svjour}
\usepackage{graphics}
\usepackage{amsmath}
\usepackage{multirow}

\begin{document}
\title{Monte Carlo integration on GPU}
%\author{First author\inst{1} \and Second author\inst{2}% etc
\author{J.~Kanzaki\inst{1}\fnmsep\thanks{e-mail: junichi.kanzaki@kek.jp}
   }                     % Do not remove
   %\institute{Insert the first address here \and the second here}
\institute{KEK, Tsukuba 305-0801, Japan}

\date{Received: date / Revised version: \today}

\abstract{
We use a graphics processing unit (GPU) for fast computations of
Monte Carlo integrations.
Two widely used Monte Carlo integration programs, VEGAS and BASES,
are parallelized on GPU.
By using $W^{+}$ plus multi-gluon production processes at LHC, we test 
integrated cross sections and execution time for programs in FORTRAN and
C on CPU and those on GPU.
Integrated results agree with each other within statistical errors.
Execution time of programs on GPU run about 50 times faster
than those in C, and more than 60 times faster than the
original FORTRAN programs.
%\PACS{
%      {PACS-key}{discribing text of that key}   \and
%      {PACS-key}{discribing text of that key}
%    } % end of PACS codes
} %end of abstract
		
\titlerunning{Fast Monte Carlo integration on GPU}
%\authorrunning{K.~Hagiwara et al.}
%

\maketitle

%%%%%%%%%%%%%% Begin Section 1 %%%%%%%%%%%%%%%%%%%%%%%%%%%%%%%%%%%%%%%%%
%\newpage
 \section{Introduction}
 \label{sec:intro}
 
  GPU (Graphics Processing Unit) is originally developed for fast 
  output of moving complex images onto computer displays.
  Because it is composed of many multi-processors, it can be used 
  as a powerful parallel processor not only for graphics
  applications but also for general purpose computations.
  GPU has already been used in scientific applications which
  require a huge number of calculations to process data as in
  astrophysics and fluid dynamics.
  Also in elementary particle physics,
  successful computations of various cross sections on GPU
  have been reported in \cite{qed-qcd-paper,giele}.
  In these studies programs on GPU were shown to run about 100 times 
  faster than those on CPU.

  Two-orders of magnitude  reduction of computation time by GPU demonstrated
  in the previous studies should greatly improve the efficiency of analysis 
  in the field of elementary particle physics.
  %  With this high performance, the application of GPU to more programs
  %  in the field of the elemetary particle physics which require
  %  large computing resources should greatly improve the
  %  the efficiency of our studies.
  %  large computing resources should greatly reduce time necessary for
  %  the computation and eventually improve the efficiency of our 
  %  computations.
  In this paper, we show that general purpose Monte Carlo
  integration programs can be adopted to run on GPU, opening the
  door of fast and economical computation to all area of research that
  makes use of Monte Carlo method.

  \section{Monte Carlo integration programs}
  \label{sec:mc}
  
  Scattering amplitudes of physics processes at LHC energies are
  expressed as complex functions of momenta and helicities of external
  particles, and kinematical distributions of produced particles are
  obtained by integrals of the squared amplitudes over many-body phase space.
  Because integration of multi-dimensional function is most conveniently 
  done with Monte Carlo integration technique,
  the method is widely used in the field of elementary particle physics.
  They are especially useful when evaluating differential 
  cross sections with experimental cuts on produced particle momenta.

   %  Cross sections of multi-particle productions in collider physics with
   %  acceptance cuts on final state particles
   %   Without this technique it is very difficult to obtain reliable cross sections
   %   of physics processes with many body final states.

   As the number of final state particles increases, the computation
   time which is necessary to obtain good accuracy of the integrated
   results grows quickly.
   It is partly because the number of sampling points during the integration 
   should be large for higher dimensional integral, and partly because
   the computation time for the scattering amplitudes also increases
   as the number of external particles grows.
   Therefore, integration of differential cross section with good accuracy
   becomes a very time consuming task for multi-particle productions
   processes.
   Significant reduction of computation time by the use of GPU will
   contribute to the improvement of the efficiency of physics analysis
   at LHC and elsewhere.
   
   %   In order to obtain good accuracy of the integrated result,
   %   the number of sampling points during the integration should be large.

   %   As the number of final states increases, more sampling points are
   %   neeeded in order to obtain good accuracy of the integrated results.
   %   And process time necessary for one function call also increases.
   %   Total resources of the execution time becomes very large.
   %   The reduction of necessary CPU time for the integration of 
   %   complex functions should be very helpful for our studies.
   %   GPU can contribute ...

   VEGAS~\cite{vegas} and its variants are widely used 
   for Monte Carlo integration.
   They are based on an iterative and adaptive Monte Carlo scheme.
   In these programs each axis of variable is divided into grids, 
   thus the integrand volume is divided into hyper cubes.
   Monte Carlo integration is performed in each hypercube and 
   variances from hypercubes are used to define new grid spacings 
   which are used in the next iteration step.
   The variance of total integral is reduced by iteration by
   iteration.
   BASES~\cite{bases} is one of its variants developed at KEK,
   which has been widely used in particle physics at colliders.
   
   In this paper we study parallelization of VEGAS and 
   BASES, by using GPU.

   %   Grid optimization ...
   %   Many types of programs have been used for the Monte Carlo integration.
   %   Among them ``VEGAS''~\cite{vegas} which adopted the alogorithm 
   %   with adaptive 
   %   multidimensional integration is most commonly used for this purpose.

   %   BASES~\cite{bases} which was developed at KEK 
   %   is another useful program
   %   for the Monte Carlo integration.
   %   Its algorithm includes ..
   %   BASES is a very powerful tool which includes not only the Monte Carlo
   %   integration package but also the event generation package, SPRING,
   %   based on the integration data.
   %   It also includes the histograming package which can be called within 
   %   the integrand function and is very useful for checking distribution of
   %   various variables.

   %   In this paper parallelization of VEGAS and BASES programs on GPU is studied.

   %%%%%%%%%%%%%% Begin Section 2 %%%%%%%%%%%%%%%%%%%%%%%%%%%%%%%%%%%%%%%%%
   \section{Parallelization of Monte Carlo integration program}
   \label{sec:parallel}

   \subsection{Program structure}
   \label{sec:parallel-structure}

   Multi-dimensional integration programs,
   VEGAS and BASES, have the following common structure:
   \begin{enumerate}
    \item initialize parameters,
    \item generate a space point within a $k$-dimensional hypercube from
	  a set of $k$ random numbers,
    \item compute an integrand function at the generated space point,
    \item accumulate function values,
    \item optimize grid spacing after accumulating $N$ function values,
    \item repeat 2-5 steps up to $M$ iterations or until the desired accuracy 
	  is reached.

   \end{enumerate}
   
   In BASES, after $M$ iterations (grid optimization phase) are done, 
   further iteration steps are executed in order to improve the accuracy
   of the integration (integration phase).
   The results of this integration phase are used for event generations
   by SPRING~\cite{bases}.

   %   \subsection{CPU time fractions}
   %   \label{sec:parallel-cpu-time}

   %   GPU has many processors and on each processor the same program is executed.
   %   Using GPU parts of program with many reputations can be parallelized.
   %   If the fraction of reputation is large, the effect of parallelization becomes 
   %   large.

   We measure fractions of CPU time for each step and find that 
   almost 98-99\% of total CPU time is used in the step 3
   where integrand function is computed.
   This fraction grows as the number of sampling points grows and 
   the complexity of the integrand function grows.
   Therefore significant reduction of total CPU time is expected by
   parallelizing function calls at all sampling points with GPU.

   \subsection{Program conversion}
   \label{sec:parallel-conversion}

   Both VEGAS and BASES are originally written in FORTRAN.
   In order to transfer function calls to GPU, they should be
   written in CUDA~\cite{cuda}, C/C++ style platform developed
   for general purpose computing on GPU.
   We first convert the FORTRAN programs into C codes.
   Then we transform the function call part further into CUDA codes.
   Due to the limited support for double precision computation
   capabilities of the GPU which we use for this study~\cite{qed-qcd-paper}, 
   floating point computations in the GPU programs are done in 
   single precision\footnote{This limitation is relaxed for NVIDIA's
   GPUs with newer architecture~\cite{fermi}.}.
   We check results and performances of all programs of FORTRAN, C and GPU
   versions.

   In the programs of GPU version, all sampling points are generated 
   on GPU and integrand function values at each space  point is computed in 
   parallel (steps 2-3).  
   Then computed function values are transferred to CPU memories.
   At the CPU side, computed function values are accumulated and grid parameters
   are optimized based on the accumulated information (steps 4-5).
   These steps are iterated and variance of integral are reduced.

   %   For the easy devlopment of programs which can be executed on GPU with 
   %   high performance NVIDIA~\cite{nvidia} introduced the software develpment
   %   package, CUDA~\cite{cuda}, with which we can write programs for GPU in
   %   C/C++.

   %%%%%%%%%%%%%% Begin Section 2 %%%%%%%%%%%%%%%%%%%%%%%%%%%%%%%%%%%%%%%%%
   \section{Computing environments}
   \label{sec:computing}

   %   Random number generation on GPU ...

   \subsection{GPU and its host PC}
   \label{sec:computing-gpu}

   We use a GeForce GTX285 by NVIDIA \cite{nvidia}
   for the computation of cross sections of physics processes
   with Monte Carlo integration.
   The GeForce GTX285 which is connected with PCI Express2$\!\times\!$16 bus
   has 30 streaming multi-processors (SM).
   Since each SM has 8 streaming processors (SP), 
   the GTX285 GPU card has 240 SP in total.
   %total 240 SP are in one GPU card.
   Other parameters of the GTX285 are summarized in Table~\ref{tab:gpu}.

   The GTX285 is controlled by Linux PC with Fedora10 (64bit)
   operating system.
   The parameters of host computer is summarized in 
   Table.~\ref{tab:host-pc}.

   In order to compile programs of GPU version,
%   For the compilation of programs in GPU version,
   we use the CUDA 
   version 2.3 toolkit  which are obtained from the NVIDIA site \cite{nvidia}.
   And for the programs in FORTRAN and C, we use gfortran and
   gcc which is automatically installed with Fedora 10.
   The version of compilers are summarized in Table.~\ref{tab:develop}.

   \begin{table}[tbh]
    \centering
    \caption{Parameters of GTX285}
    \label{tab:gpu}       % Give a unique label
    \smallskip
    \begin{tabular}{|c|c|}
     \hline
     % & GTX285
     %     % &  TESLA
     % \\
     % \hline\hline
     Number of      & \multirow{2}{*}{30} \\
     multiprocessor &                     \\  \hline
     Number of core & 240   \\   \hline
     Total amount of    & \multirow{2}{*}{2GB} \\
     global memory &                             \\  \hline
     Total amount of      & \multirow{2}{*}{64kB}  \\
     constant memory &   	    \\    \hline
     Total amount of shared & \multirow{2}{*}{16kB}    \\
     memory per block   &   \\    \hline
     Total number of registers & \multirow{2}{*}{16kB}   \\
     available per bloc  &  \\  \hline
     Clock rate & 1.48GHz   \\    \hline
    \end{tabular}
   % \end{table}
    % \begin{table}[tbh]
    \centering
    \caption{Host PC environment}
    \label{tab:host-pc}       % Give a unique label
    \begin{tabular}{|c|c|} \hline
     CPU & Core i7 2.67GHz \\ \hline
     L2 Cache & 8MB  \\ \hline
     Memory & 6GB  \\ \hline
     Bus Speed & 1.333GHz  \\ \hline
     OS & Fedora 10 (64 bit)  \\ \hline
     % \noalign{\smallskip}
    \end{tabular}
      % \end{table}
    % \begin{table}[tbh]
    \centering
    \caption{development environment}
    \label{tab:develop}       % Give a unique label
    \begin{tabular}{|c|c|} \hline
     nvcc         & Rel. 2.3 (V0.2.1221)  \\ \hline
     CUDA Driver  & Ver.2.30 \\ \hline
     CUDA Runtime & Ver 2.30 \\ \hline
     \hline
     gcc          & 4.3.2 (Red Hat 4.3.2-7) \\ \hline
     gfortran     & 4.3.2 (Red Hat 4.3.2-7) \\ \hline
     % \noalign{\smallskip}
    \end{tabular}
      \end{table}

      \subsection{Process time measurement}
      \label{sec:results-time-measurement}

      For comparisons of execution time, we measure the time
      between the start of VEGAS/BASES programs and the end of them,
      i.e. between the step 1 and the completion of the step 6,
      including the steps 4 and 5 that are processed on CPU.
      For FORTRAN programs, an intrinsic procedure of gfortran,
      ``\texttt{cpu\_time}'',
      is used for the measurement of elapsed CPU time.
      For C and GPU programs, a system call,
      ``\texttt{getrusage}'',  is used for the time measurements.

      \section{Physics process}
      \label{sec:process}

      In order to test the GPU version of VEGAS and BASES, called
      gVEGAS and gBASES\footnote{Sample source codes of gVEGAS are
      available on the web page: 
      \texttt{http://madgraph.kek.jp/KEK/GPU/gVEGAS/example/}}
      respectively, we compare total cross sections of multi-particle
      production process at LHC.
      In particular, we report results on the following processes 
      \begin{equation}
       u\overline{d}\rightarrow W^{+}(\rightarrow \mu^{+}\nu_{\mu})
	+n\textrm{ gluons } (n=0\sim 4)
	\label{eq:process}
      \end{equation}
      with semi-realistic final state cuts at LHC.
      The dimension of integral is $3(n\!+\!2)\!-\!4$ from the phase space,
      2 from the parton distributions (PDF), and 1 for the helicity 
      summation, and hence $3n\!+\!5$; hence from 5-dimensional integral
      for no gluon ($n\!=\!0$) to 17-dimensional integral for 4
      gluons ($n\!=\!4$).

      The degree of the complexity (length) of the integral function
      can be estimated from the number of contributing Feynman 
      diagrams and the number of independent color-basis vectors
      as listed in Table~\ref{tab:sample}.
      Previous studies \cite{qed-qcd-paper} show that the performance of
      GPU computation is limited by the product of these two numbers,
      the processes eq.~(\ref{eq:process}) cover program size of four orders
      of magnitude difference.

      In order to simulate realistic LHC experiments, We introduce the
      following final state cuts. 
      For gluons,
      \begin{subequations}
       \begin{eqnarray}
	%     {\renewcommand\arraystretch{1.2}
	 |\eta_{i}| &<& 5, \label{eq:cuts-eta} \\
	p_{\mathrm{T}i} &>& 20\,\mathrm{GeV}, \label{eq:cuts-pt} \\
	p_{\mathrm{T}ij} &>& 20\,\mathrm{GeV}, \label{eq:cuts-ptjj}
	 %      }
       \end{eqnarray}
       \label{eq:cuts}
      \end{subequations}
      where $\eta_{i}$ and $p_{\mathrm{T}i}$ are the rapidity and the
      transverse momentum of the $i$-th jet, respectively, in the $pp$
      collisions rest frame along the right-moving ($p_{\mathrm{z}}\!=\!|p|$)
      proton momentum direction, and $p_{\mathrm{T}ij}$ is the relative
      transverse momentum~\cite{durham-jet} between the jets $i$ and $j$
      defined by
      \begin{subequations}
       \begin{eqnarray}
	%    {\renewcommand\arraystretch{2.}
	 p_{\mathrm{T}ij} & \equiv &
	 \min(p_{\mathrm{T}i},p_{\mathrm{T}j})\,\Delta R_{ij}, \\ 
	\Delta R_{ij} & = &
	 \sqrt{\Delta\eta_{ij}^{2}+\Delta\phi_{ij}^{2}}.
       \end{eqnarray}
       \label{eq:isolationcuts}
      \end{subequations}
      Here $\Delta R_{ij}$ measures the boost-invariant angular separation
      between jets. 
      For $\mu^{+}$ from $W^{+}$ decay, we require
      \begin{subequations}
       \begin{eqnarray}
	|\eta_{l}| & < & 2.5,
	 \label{eq:lcuts-eta} \\
	p_{\mathrm{T}l} & > & 20\,\mathrm{GeV}
	 \label{eq:lcuts-pt} 
       \end{eqnarray}
       \label{eq:leptoncuts}
      \end{subequations}

      As for the parton distribution function (PDF), we use the set
      CTEQ6L1~\cite{cteq}  and the factorization scale is chosen to be 
      the $Z$ boson mass.
      The QCD coupling constant is also fixed as
      $\alpha_s(m_{Z})_{\overline{\mathrm{MS}}}\! =\!0.118$~\cite{pdg}.

      For the computation of helicity amplitudes of these processes, 
      HELAS~\cite{helas} for FORTRAN programs and its C/GPU version, 
      HEGET~\cite{qed-qcd-paper} are used.

      \begin{table}
       \centering
       \caption{$u\overline{d}\rightarrow W^{+}(\rightarrow \mu^{+}\nu_{\mu})
       +\textrm{gluons}$}
       \label{tab:sample}       % Give a unique label
       \begin{tabular}{|c|c|c|} \hline
	Number of & Number of & Number of \\ 
	gluons    & diagrams  & color bases \\ \hline
	0 & 1 & 1 \\
	1 & 2 & 1 \\
	2 & 8 & 2 \\
	3 & 54 & 6 \\
	4 & 516 & 24 \\ \hline
       \end{tabular}
      \end{table}

      \section{Results}
      \label{sec:results}

      \subsection{Parameters of the integration programs}
      \label{sec:results-parameters}

      In order to control the behavior of the Monte Carlo integration 
      by VEGAS and BASES, user can give them the following parameters:
      \begin{itemize}
       \item number of total function calls in one iteration 
	     step (\texttt{NCALL}),
       \item number of maximum iteration steps (\texttt{ITMX}), and
       \item desired accuracy of the integration (\texttt{ACC}).
      \end{itemize}
      \texttt{NCALL} is the number $N$ in step 5 and \texttt{ITMX}
      is the number $M$ of the step 6 in Section~\ref{sec:parallel-structure}.
      Iteration steps of BASES are separated into two phases: 
      the grid optimization step and the integration step.
      Accordingly, \texttt{ITMX} and \texttt{ACC} are also separated as:
      \begin{itemize}
       \item number of maximum iteration steps (\texttt{ITMX1}), and
       \item desired accuracy of integration (\texttt{ACC1})
      \end{itemize}
      for the grid optimization phase, and
      \begin{itemize}
       \item number of maximum iteration steps (\texttt{ITMX2}), and
       \item desired accuracy of integration (\texttt{ACC2})
      \end{itemize}
      for the integration phase.

      Parameter values used in this study are summarized
      in Table~\ref{tab:parameter}.
      \begin{table}
       \centering
       \caption{Parameters for integrations}
       \label{tab:parameter}       % Give a unique label
       \begin{tabular}{|c|c|c|c|c|} \hline
	Number of & \multirow{2}{*}{\texttt{NCALL}} &
	\multirow{2}{*}{\texttt{ITMX}} & \multirow{2}{*}{\texttt{ITMX1}} &
	\multirow{2}{*}{\texttt{ITMX2}}\\ 
	gluons  & & & & \\ \hline
	0 & $10^{6}$ & 10 & 5 & 5 \\
	1 & $10^{6}$ & 10 & 5 & 5 \\
	2 & $10^{6}$ & 10 & 5 & 5 \\
	3 & $10^{7}$ & 10 & 5 & 5 \\
	4 & $10^{7}$ & 10 & 5 & 5 \\ \hline
       \end{tabular}
      \end{table}
      In order to keep the total amount of computations to be the same among
      all the programs, all desired accuracies, \texttt{ACC} for VEGAS and 
      \texttt{ACC1} and \texttt{ACC2}, are set to an extremely small value 
      (0.001\%) which cannot be reached by MC sampling of
      $\texttt{NCALL}\!\times\!\texttt{ITMX}$ points used in this study: 
      see Table~\ref{tab:parameter}.
      For BASES, numbers
      of iteration steps for the grid optimization and integration phases 
      are set to be equal ($\texttt{ITMX1}\! =\! \texttt{ITMX2}$), and 
      their sum is set the same as \texttt{ITMX} of VEGAS programs 
      ($\texttt{ITMX1}\!+\!\texttt{ITMX2}\!=\!\texttt{ITMX}$).
      In summary, we accumulate $10^{7}$ sample points for processes 
      up to two gluons ($n=0, 1, 2$) and $10^{8}$ points for those with
      more gluons ($n=3$ and 4).
      \subsection{Total cross section computation}
      \label{sec:results-sigma}
      
      Total cross sections for processes in eq.~(\ref{eq:process}) 
      with experimental cuts (eqs.~\ref{eq:cuts}-\ref{eq:leptoncuts}) 
      are listed in Table~\ref{tab:sigma}.
      They are computed with programs in FORTRAN, C and CUDA (GPU).
      Cross sections from different of programs agree to each other
      within their statistical errors.
      In addition, they agree with the results from the event generator
      MadGraph/MadEvent~\cite{madgraph,madevent,newmad}.
      \begin{table*}[tbh]
       \centering
       \begin{tabular}{|c|c|c|c|c|c|c|c|c|} \hline
	No. of & \multicolumn{3}{|c|}{VEGAS} & \multicolumn{3}{|c|}{BASES} &
	\multirow{2}{*}{MG/ME} & \multirow{2}{*}{[fb]}\\ \cline{2-7}
	gluons & FORTRAN & C & GPU & FORTRAN & C & GPU &  & \\ \hline
	0 &
	$2.137\!\pm\! 0.001$ & $2.138\!\pm\! 0.001$ & $2.137\!\pm\! 0.001$ &
	$2.137\!\pm\! 0.001$ & $2.137\!\pm\! 0.001$ & $2.137\!\pm\! 0.001$ &
	$2.138\!\pm\! 0.002$ & $\times 10^{6} $ \\ \hline
	1 & 
	$1.783\!\pm\! 0.001$ & $1.783\!\pm\! 0.001$ & $1.780\!\pm\! 0.001$ &
	$1.785\!\pm\! 0.001$ & $1.784\!\pm\! 0.001$ & $1.782\!\pm\! 0.001$ &
	$1.773\!\pm\! 0.003$ & $\times 10^{5}$ \\ \hline
	2 &
	$1.873\!\pm\! 0.007$ & $1.853\!\pm\! 0.006$ & $1.843\!\pm\! 0.006$ &
	$1.876\!\pm\! 0.007$ & $1.883\!\pm\! 0.010$ & $1.870\!\pm\! 0.007$ &
	$1.874\!\pm\! 0.002$ & $\times 10^{4}$ \\ \hline
	3 & 
	$2.868\!\pm\! 0.008$ & $2.881\!\pm\! 0.009$ & $2.832\!\pm\! 0.010$ &
	$2.860\!\pm\! 0.010$ & $2.855\!\pm\! 0.014$ & $2.907\!\pm\! 0.012$ &
	$2.845\!\pm\! 0.005$ & $\times 10^{3}$ \\ \hline
	4 &
	$6.186\!\pm\! 0.041$ & $6.054\!\pm\! 0.081$ & $6.157\!\pm\! 0.073$ & 
	$6.078\!\pm\! 0.134$& $6.191\!\pm\! 0.068$ & $6.385\!\pm\! 0.235$ &
	$6.070\!\pm\! 0.010$ & $\times 10^{2}$ \\ \hline
       \end{tabular}
       \caption{Total cross sections of $u\overline{d}\!\rightarrow
       \!W^{+}(\!\rightarrow\!\mu^{+}\nu_{\mu})+n\textrm{-gluons}$} 
       computed by programs in FORTRAN, C, CUDA (GPU) and MadGraph/MadEvent.
       \label{tab:sigma}
      \end{table*}

      \subsection{Parameters of the kernel program}
      \label{sec:results-kernels}

      The performance of GPU programs largely depends on
      parameters of kernel programs executed on GPU.
      Most significant parameters which affect the process time of programs 
      are:
      \begin{itemize}
       \item number of registers allocated to a thread, and
       \item number of threads in a thread block.
      \end{itemize}
      Details of kernel parameters are explained in~\cite{qed-qcd-paper}.
      In this study we use 64 as a number of registers allocated a
      thread and 256 as a number of threads in a thread block.
      From the detailed study of dependence of performance on
      these parameters we find that they give almost the best
      performance for all processes in this study.
      
      Number of thread blocks in a grid (= a set of thread blocks),
      which is executed with a single kernel call, are set to be
      equal to \texttt{NCALL}, so that one iteration of Monte Carlo 
      integration steps is executed by a single kernel call.

      \subsection{Process time comparisons}
      \label{sec:results-time}

      In Table~\ref{tab:time} measured process time for a single function call 
      is listed for all programs.
      As explained above, the process time per single function call is
      obtained by dividing the total computation time by $10^{7}$ for processes
      with up to two gluons ($n\!=\!0,1,2$) and by $10^{8}$ for those
      with more gluons ($n\!=\!3$ and 4).

      Numbers in parentheses in the FORTRAN and C columns in Table~\ref{tab:time}
      are the ratio of the process time as compared to that of GPU.
      About a factor of 50 times more sampling is possible with GPU
      as compared to the C  programs on CPU.
      During the comparison of process time,
      we find that the original FORTRAN codes run slower than the C-version.
      Because the total process time for these CPU programs is dominated
      by the function (amplitude) computation, this FORTRAN-to-C
      ratio can be originated from the difference of handling complex 
      numbers which appears in amplitude computations.
      We use in-line functions for the computations of complex numbers in C,
      which might have better efficiency compared with built-in 
      complex functions in FORTRAN.
%      In FORTRAN programs, we use built-in COMPLEX type, whereas 
%      we use in-line functions which handle a structure with
%      two real numbers for the computation of complex numbers in 
%      C programs.

%      This difference between C and FORTRAN versions of programs depends
%      on the nature of integrand functions.
%      The FORTRAN-to-C ratio can be as large as a factor of time
%      for the simples process with no-gluon emissions ($n=0$),
%      suggesting inefficiency in process other than the function
%      (amplitude) computation.

\begin{figure}[htb]
       \centering
       \resizebox{0.45\textwidth}{!}{%
       \includegraphics{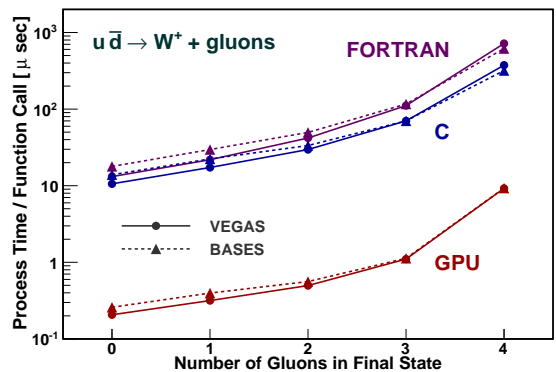}
       }
       \caption{Process time of a single function call for $u\overline{d}\!
       \rightarrow\! W^{+}(\rightarrow\!\mu^{+}\nu_{\mu})+n\textrm{-gluons}$.}
       \label{fig:time}       % Give a unique label
      \end{figure}

\begin{figure}[htb]
       \centering
       \resizebox{0.45\textwidth}{!}{%
       \includegraphics{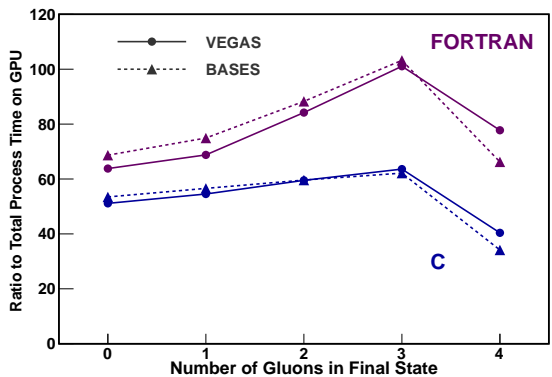}
       }
       \caption{Process time ratios of FORTRAN and C programs to the 
       corresponding GPU program}
       \label{fig:ratio}       % Give a unique label
      \end{figure}

      In Fig.~\ref{fig:time} process time for a single function call is plotted
      against the number of gluons in the final state.
      And in Fig.~\ref{fig:ratio} ratios of process time between programs on
      CPU (FORTRAN/C) and those on GPU are plotted.
      Differences of process time between VEGAS and BASES are small.
%      Programs in C show better performance than those in FORTRAN
%      which are executed on the same CPU.
      Programs which are executed on GPU 
      can run about 50 times faster than those in C.
      Compared with original FORTRAN version programs, the differences
      of performance become larger.

      When the final state has 4 gluons, the size of GPU program becomes
      large and requires more access to local memories.
      From previous studies on performances of 
      GPU programs~\cite{qed-qcd-paper},
      larger programs show worse performance on execution time.
      Still the VEGAS(BASES) programs for the 4 gluon production process
      runs 40 (34) times faster on GPU than the C-program runs on CPU.
      
      \begin{table*}[tbh]
       \centering
       \begin{tabular}{|c|c|c|c|c|c|c|} \hline
	\multirow{2}{*}{No. of gluons} & 
	\multicolumn{3}{|c|}{VEGAS [$\mu$sec]} &
	\multicolumn{3}{|c|}{BASES [$\mu$sec]} \\ \cline{2-7}
	& FORTRAN & C & GPU & FORTRAN & C & GPU \\ \hline
	0 &
	1.32 (63.8) & 1.06 (51.2) & 0.0207 &
	1.78 (68.7) & 1.39 (53.5) & 0.0260 \\
	1 &
	2.19 (68.8) & 1.73 (54.6) & 0.0318 &
	2.97 (75.0) & 2.24 (56.6) & 0.0396 \\ 
	2 &
	4.19 (84.2) & 2.96 (59.5) & 0.0497 &
	4.97 (88.3) & 3.35 (59.6) & 0.0563 \\
	3 &
	11.1 (101) & 7.00 (63.6) & 0.110 &
	11.7 (103) & 7.02 (62.2) & 0.113 \\
	4 &
	72.1 (77.8) & 37.4 (40.4) & 0.927 &
	61.6 (66.2) & 31.8 (34.2) & 0.931 \\ \hline
       \end{tabular}
       \caption{Process time for a single function call in VEGAS and BASES
       on CPU with FORTRAN or C, and on GPU with CUDA.  Numbers in the
       parentheses of the FORTRAN and C columns are the ratio of process
       time relative to that of GPU.}
       \label{tab:time}
      \end{table*}

      \section{Summary}
      \label{sec:summary}

      Based on VEGAS and BASES programs written in FORTRAN, we have 
      developed Monte Carlo integration programs, gVEGAS and gBASES
      respectively, which can be executed on
      NVIDIA's GPU using the CUDA development kit.
      We have tested their performance with the computation of total
      cross sections of processes, $u\overline{d}\!\rightarrow\!
      W^{+}(\!\rightarrow\!\mu^{+}\nu_{\mu})
      +n\textrm{-gluons } (n\!=\!0\!\sim\! 4)$, in $pp$ collisions at
      $\sqrt{s}=14\mathrm{TeV}$.
      Total cross sections agree with each other
      within statistical errors for all programs.
      Both VEGAS and BASES programs on GPU run about 50 times faster 
      than the same  programs written in C,  which are converted from 
      the original FORTRAN version programs.
      Compared with FORTRAN programs their GPU version programs show
      more than 60 times better performance in execution time.
      For the process with 4 gluons, the size of GPU programs becomes 
      large and their relative performance become worse than small programs.
%      Even for this process, BASES program can be executed about 30 times
%      faster on GPU than that of C-version on CPU.

      \begin{acknowledgement}{\textit{Acknowledgment}.}
       The author wishes to thank Kaoru Hagiwara for
       his encouragement and advice throughout this work.
       He also thanks Naotoshi Okamura in valuable discussions.
       This work is supported in part by the Grant-in-Aid for Scientific
       Research from the Japan Society 
       for the Promotion of Science (No. 20340064).
      \end{acknowledgement}
      
      %%%%%%%%%%%%%% Begin Appendix %%%%%%%%%%%%%%%%%%%%%%%%%%%%%%%%%%%%%%%%%
%      \newpage
\appendix
\def\thesection{Appendix \Alph{section}}

 \section{Sample codes for gVEGAS}
	     
	     Sample source codes of the gVEGAS program are available
	     from the web page:\\
	     \texttt{http://madgraph.kek.jp/KEK/GPU/gVEGAS/example/}.

	     \noindent
 	     They include a minimum set of source files which are necessary 
	     to do Monte Carlo integration with the VEGAS algorithm on GPU,
 	     but do not include
	     \texttt{Makefile} which largely depends on user's environment
	     of development.
	    
    \subsection{User programs}

    Sample codes include two user programs: \texttt{gVegasMain.cu} and
    \texttt{gVegasFunc}.
    They should be customized by user to the task one intends to 
    perform.

  \subsubsection{\texttt{gVegasMain.cu}}
	     
 	     \texttt{gVegasMain.cu} includes a sample main program for
	     Monte Carlo integration where user can 
	     set parameters for the integration. 
	     Typical parameters are:
	     \[
   	     \begin{array}{ll}
	      \texttt{nBlockSize} &
	       \textrm{size of a thread block of a kernel program} \\
	      & \textrm{on GPU} \\
	      \texttt{ndim} &
	       \textrm{number of independent variables of integrand} \\
	       & \textrm{function} \\
	      \texttt{ncall} &
	       \textrm{number of sample points per iteration} \\
	      \texttt{itmx} &
	       \textrm{maximum number of iterations} \\
	      \texttt{acc} &
	       \textrm{required accuracy during iterations}
	     \end{array}
	     \]
	     All these parameters can be set within \texttt{gVegasMain.cu}.

  \subsubsection{\texttt{gVegasFunc.cu}}
	     
      	     User function program integrated in the program is described
	     in \texttt{gVegasFunc.cu}.
	     The calling sequence of user functions is
	     \[
	     \texttt{float func(float* rx, float wgt)}
	     \]
	     where \texttt{rx} includes a set of variables and 
	     \texttt{wgt} is a function weight.

  \subsection{Internal programs}

  The gVEGAS consists of the following programs which are
  included in the sample codes:
	      \[
   	     \begin{array}{ll}
	      \texttt{gVegas.cu} & 
	       \textrm{main program of \texttt{gVEGAS} system} \\
	      \lefteqn{\texttt{gVegasCallFunction.cu}} \\
	      & \textrm{kernel program which runs on GPU called} \\
	      & \textrm{from \texttt{gVegas.cu}.} \\
	      \texttt{xorshift.cu} & 
	       \textrm{random number generator on GPU}
	     \end{array}
	     \]

  \subsection{Header files}

  The following header files which are necessary for the gVEGAS system
  are also included in the sample codes:
	     \[
   	     \begin{array}{ll}
	      \texttt{gvegas.h} &
	       \textrm{includes \texttt{nBlockSize} which user} \\
	      & \textrm{can set in \texttt{gVegasMain.cu} } \\
	      \texttt{vegasconst.h} &
	       \textrm{includes internal constants which} \\
	      & \textrm{are located at constant memory of} \\
	       & \textrm{GPU} \\
	      \texttt{vegas.h} &
	       \textrm{includes internal \texttt{gVEGAS} parameters} \\
	      \texttt{kernels.h} &
	       \textrm{a list of kernel programs which are} \\
	      & \textrm{included at CUDA compilation.} 
	     \end{array}
	     \]

 %%%%%%%%%%%%%% Begin Reference List %%%%%%%%%%%%%%%%%%%%%%%%%%%%%%%%%%%%%%%%%
 % 
 % BibTeX users please use
 % \bibliographystyle{}
 % \bibliography{}
 % \bibliographystyle{phjcp}
 % \bibliography{reference}

\end{document}